\documentclass[twocolumn,showpacs,aps,prl,superscriptaddress,letterpaper]{revtex4}
\usepackage{graphicx}
\usepackage{dcolumn}
\usepackage{amsmath}
\usepackage{epsfig}

\def\ppm{\ensuremath{\pm}}
\def\spk{\ensuremath{S_{\phi K}}\xspace}
\def\cpk{\ensuremath{C_{\phi K}}\xspace}
\def\spkp{\ensuremath{S_{\phi K^+}}\xspace}
\def\cpkp{\ensuremath{C_{\phi K^+}}\xspace}

\def\Bflav {\ensuremath{B_{\text{flav}}}\xspace}
\def\Btag {\ensuremath{B_{\text{tag}}}\xspace}

\long\def\inst#1{\par\nobreak\kern 4pt\nobreak
    {\itshape #1}\par\vskip 10pt plus 3pt minus 3pt}

\RequirePackage{xspace}

\usepackage{relsize}
\def\babar{\mbox{\slshape B\kern-0.1em{\smaller A}\kern-0.1em
    B\kern-0.1em{\smaller A\kern-0.2em R}}}

\def\epem       {\ensuremath{e^+e^-}\xspace}

\def\pip   {\ensuremath{\pi^+}\xspace}
\def\pim   {\ensuremath{\pi^-}\xspace}

\def\Kbar  {\kern 0.2em\overline{\kern -0.2em K}{}\xspace}

\def\Kz    {\ensuremath{K^0}\xspace}
\def\Kzb   {\ensuremath{\Kbar^0}\xspace}
\def\KzKzb {\ensuremath{\Kz \kern -0.16em \Kzb}\xspace}
\def\Kp    {\ensuremath{K^+}\xspace}
\def\Km    {\ensuremath{K^-}\xspace}

\def\KpKm  {\ensuremath{\Kp \kern -0.16em \Km}\xspace}
\def\KS    {\ensuremath{K^0_{\scriptscriptstyle S}}\xspace} 
\def\KL    {\ensuremath{K^0_{\scriptscriptstyle L}}\xspace}

\def\Dbar    {\kern 0.2em\overline{\kern -0.2em D}{}\xspace}

\def\Dz      {\ensuremath{D^0}\xspace}
\def\Dzb     {\ensuremath{\Dbar^0}\xspace}
\def\DzDzb   {\ensuremath{\Dz {\kern -0.16em \Dzb}}\xspace}
\def\Dp      {\ensuremath{D^+}\xspace}
\def\Dm      {\ensuremath{D^-}\xspace}

\def\DpDm    {\ensuremath{\Dp {\kern -0.16em \Dm}}\xspace}

\def\Bbar    {\kern 0.18em\overline{\kern -0.18em B}{}\xspace}

\def\Bz      {\ensuremath{B^0}\xspace}
\def\Bzb     {\ensuremath{\Bbar^0}\xspace}
\def\BzBzb   {\ensuremath{\Bz {\kern -0.16em \Bzb}}\xspace}
\def\Bu      {\ensuremath{B^+}\xspace}
\def\Bub     {\ensuremath{B^-}\xspace}

\def\BpBm    {\ensuremath{\Bu {\kern -0.16em \Bub}}\xspace}

\def\BorBbar    {\kern 0.18em\optbar{\kern -0.18em B}{}\xspace}
\def\DorDbar    {\kern 0.18em\optbar{\kern -0.18em D}{}\xspace}
\def\KorKbar    {\kern 0.18em\optbar{\kern -0.18em K}{}\xspace}

\def\jpsi     {\ensuremath{{J\mskip -3mu/\mskip -2mu\psi\mskip 2mu}}\xspace}

\mathchardef\Upsilon="7107
\def\Y#1S{\ensuremath{\Upsilon{(#1S)}}\xspace}

\mathchardef\Deltares="7101
\mathchardef\Xi="7104
\mathchardef\Lambda="7103
\mathchardef\Sigma="7106
\mathchardef\Omega="710A

\def\Deltabar{\kern 0.25em\overline{\kern -0.25em \Deltares}{}\xspace}
\def\Lbar{\kern 0.2em\overline{\kern -0.2em\Lambda\kern 0.05em}\kern-0.05em{}\xspace}
\def\Sigbar{\kern 0.2em\overline{\kern -0.2em \Sigma}{}\xspace}
\def\Xibar{\kern 0.2em\overline{\kern -0.2em \Xi}{}\xspace}
\def\Obar{\kern 0.2em\overline{\kern -0.2em \Omega}{}\xspace}
\def\Nbar{\kern 0.2em\overline{\kern -0.2em N}{}\xspace}
\def\Xb{\kern 0.2em\overline{\kern -0.2em X}{}\xspace}

\def\pt         {\mbox{$p_T$}\xspace}
\def\mes        {\mbox{$m_{\rm ES}$}\xspace}

\def\DeltaE     {\mbox{$\Delta E$}\xspace}

\newcommand{\tev}{\ensuremath{\mathrm{\,Te\kern -0.1em V}}\xspace}
\newcommand{\gev}{\ensuremath{\mathrm{\,Ge\kern -0.1em V}}\xspace}
\newcommand{\mev}{\ensuremath{\mathrm{\,Me\kern -0.1em V}}\xspace}
\newcommand{\kev}{\ensuremath{\mathrm{\,ke\kern -0.1em V}}\xspace}
\newcommand{\ev}{\ensuremath{\mathrm{\,e\kern -0.1em V}}\xspace}
\newcommand{\gevc}{\ensuremath{{\mathrm{\,Ge\kern -0.1em V\!/}c}}\xspace}
\newcommand{\mevc}{\ensuremath{{\mathrm{\,Me\kern -0.1em V\!/}c}}\xspace}
\newcommand{\gevcc}{\ensuremath{{\mathrm{\,Ge\kern -0.1em V\!/}c^2}}\xspace}
\newcommand{\mevcc}{\ensuremath{{\mathrm{\,Me\kern -0.1em V\!/}c^2}}\xspace}

\def\mum  {\ensuremath{\,\mu\rm m}\xspace}

\def\mus  {\ensuremath{\rm \,\mus}\xspace}

\def\mus        {\ensuremath{\,\mu{\rm s}}\xspace}    

\def\to                 {\ensuremath{\rightarrow}\xspace}

\def\pep2{PEP-II}

\newcommand{\dedx}{\ensuremath{\mathrm{d}\hspace{-0.1em}E/\mathrm{d}x}\xspace}

\def\gsim{{~\raise.15em\hbox{$>$}\kern-.85em
          \lower.35em\hbox{$\sim$}~}\xspace}
\def\lsim{{~\raise.15em\hbox{$<$}\kern-.85em
          \lower.35em\hbox{$\sim$}~}\xspace}

\def\CP                {\ensuremath{C\!P}\xspace}

\def\stwob{\ensuremath{\sin\! 2 \beta   }\xspace}

\def\deltaz{\ensuremath{{\rm \Delta}z}\xspace}
\def\deltat{\ensuremath{{\rm \Delta}t}\xspace}
\def\deltamd{\ensuremath{{\rm \Delta}m_d}\xspace}

\xspace

\newcommand{\jprlBase}       {Phys.\ Rev.\ Lett.\xspace}
\newcommand{\jprBase}        {Phys.\ Rev.\xspace}
\newcommand{\jplBase}        {Phys.\ Lett.\xspace}
\newcommand{\nimBaseA}       {Nucl.\ Instr.\ Meth.\xspace}

\newcommand{\npBase}         {Nucl.\ Phys.\xspace}

\newcommand{\nima}      [1]  {\nimBaseA~A~{\bf #1}}

\newcommand{\npb}       [1]  {\npBase\ B~{\bf #1}}

\newcommand{\plb}       [1]  {\jplBase\ B~{\bf #1}}

\newcommand{\jprl}      [1]  {\jprlBase\ {\bf #1}}
\newcommand{\jprd}      [1]  {\jprBase\ D~{\bf #1}}

\newcommand{\progtp}    [1]  {{Prog.\ Theor.\ Phys.\ {\bf #1}}}

\def\jetset74   {\mbox{\tt Jetset \hspace{-0.5em}7.\hspace{-0.2em}4}\xspace}

\newcommand{\BABARPubYear}    {04}
\newcommand{\BABARPubNumber}  {004}

\newcommand{\SLACPubNumber} {10382}

\begin{document}
\begin{flushleft}
\babar-PUB-\BABARPubYear/\BABARPubNumber\\
SLAC-PUB-\SLACPubNumber\\
[10mm]
\end{flushleft}

\title{
\large \bfseries Measurement of the Time-Dependent \emph{CP} Asymmetry 
in the \boldmath{$B^0 \to \phi K^0$} Decay
}

%
\author{B.~Aubert}
\author{R.~Barate}
\author{D.~Boutigny}
\author{F.~Couderc}
\author{J.-M.~Gaillard}
\author{A.~Hicheur}
\author{Y.~Karyotakis}
\author{J.~P.~Lees}
\author{V.~Tisserand}
\author{A.~Zghiche}
\affiliation{Laboratoire de Physique des Particules, F-74941 Annecy-le-Vieux, France }
% ...
\collaboration{The \babar\ Collaboration}
\noaffiliation

\date{\today}


\begin{abstract}
We present a measurement of the time-dependent \CP asymmetry for the neutral
$B$-meson decay $\Bz\to\phi K^0$. We use a sample of approximately 114 million $B$-meson 
pairs taken at the $\Upsilon(4S)$ resonance with the \babar\ detector at the \pep2\ 
$B$-meson Factory at SLAC.
We reconstruct the \CP eigenstates $\phi \KS$ and $\phi \KL$
where $\phi\to K^+K^-$, $\KS\to\pi^+\pi^-$, and \KL\ is observed via its
hadronic interactions. The other $B$ meson in the event
is tagged as either a $\Bz$ or $\Bzb$ from its decay products.
The values of the \CP-violation parameters 
are $\spk = 0.47 \pm 0.34\mbox{(stat)} ^{+0.08}_{-0.06}\mbox{(syst)}$
and $\cpk = 0.01 \pm 0.33\mbox{(stat)} \pm 0.10\mbox{(syst)}$.  
\end{abstract}


\maketitle

Decays of $B$ mesons into charmless hadronic final states with a 
$\phi$ meson are dominated by $b\to s\bar{s}s$ gluonic penguin
amplitudes, possibly with  
smaller contributions from electroweak penguins, while other
Standard Model (SM) amplitudes are strongly suppressed~\cite{one}.
In the SM, \CP violation arises from a single complex phase in the
Cabibbo--Kobayashi--Maskawa (CKM) quark-mixing matrix~\cite{ckm}.
Neglecting CKM-suppressed contributions, the time-dependent 
\CP-violating asymmetries in the decays $\Bz\to \phi K^0$ and  
$\Bz\to \jpsi K^0$ are proportional to the same parameter 
$\sin 2\beta$~\cite{grossman}, where the latter decay is dominated by 
tree diagrams.
Since many scenarios of physics beyond the SM introduce additional 
diagrams with heavy particles in the penguin loops and new \CP-violating 
phases, comparison of \CP-violating observables with SM expectations is 
a sensitive probe for new physics. 
Measurements of \stwob in $B$ decays to charmonium such as $\Bz\to J/\psi \KS$ 
have been reported by the \babar~\cite{sin2bnewbabar} and Belle~\cite{sin2bnewbelle}
collaborations, and the world average for \stwob is $0.731\pm 0.056$~\cite{pdg}.
The Belle collaboration measures $\stwob = -0.96\pm 0.50^{+0.09}_{-0.11}$ in the
decay $B^0\to\phi \KS$~\cite{fu}.

In this letter we report a measurement of the time-dependent \CP\ asymmetry
in the final state $\phi K^0$
based on an integrated luminosity of approximately 108~fb$^{-1}$
collected at the $\Upsilon(4S)$ resonance 
with the \babar\ detector~\cite{Aubert:2001tu} at the \pep2\ asymmetric \epem 
collider~\cite{pep} located at the Stanford Linear Accelerator Center. 

From a $\Bz\Bzb$ meson pair we fully reconstruct one meson, $B_{CP}$, in the final
state $\phi K^0$, and partially reconstruct the recoil $B$ meson, \Btag. 
We examine \Btag for evidence that it decayed either as \Bz or \Bzb (flavor tag). 
The asymmetric beam configuration in the laboratory frame provides a boost
of $\beta\gamma = 0.56$ to the $\Upsilon(4S)$, which allows the determination
of the proper decay time difference $\Delta t = t_{CP} - t_{\text{tag}}$
from the vertex separation of the two neutral $B$ mesons along the beam ($z$) axis.
The decay rate ${\text{f}}_+({\text{f}}_-)$ when the tagging meson is a $\Bz (\Bzb)$ 
is given by 
\begin{eqnarray}
{\text{f}}_\pm(\, \deltat)& = &{\frac{{\text{e}}^{{- \left| \deltat 
\right|}/\tau_{\Bz} }}{4\tau_{\Bz}}}  \, [
\ 1 \hbox to 0cm{} \pm \,\spk \sin{( \deltamd  \deltat )} \nonumber \\
& & \mp \,\cpk \cos{( \deltamd  \deltat) }   ], 
\label{eq:timedist}
\end{eqnarray}
where $\tau_{\Bz}$ is the neutral $B$ meson mean lifetime, 
and \deltamd is the \Bz--\Bzb oscillation frequency.
The time-dependent \CP -violating asymmetry is defined as
$A_{\CP} \equiv ({\text{f}}_+  -  {\text{f}}_- )/
({\text{f}}_+ + {\text{f}}_- )$.
In the SM, decays that proceed purely via the $b\to s\bar{s}s$ penguin transitions have 
\CP parameters $\spk = -\eta_f \sin 2\beta$ and $\cpk = 0$, where $\beta \equiv \text{arg} 
\left [ -V_{cd}^{} V_{cb}^\ast / V_{td}^{} V_{tb}^\ast\right]$.
Here $V_{ik}$ is the CKM matrix element for quarks $i$ and $k$, and
the \CP eigenvalue is $\eta_f=-1$ ($+1$) for $\phi\KS$ ($\phi\KL$).

The $B_{CP}$ candidate is reconstructed in the decay mode
$\phi K^0$ with $\phi\rightarrow K^+K^-$; the $K^0$ is either 
a \KL\ or a $\KS\rightarrow\pi^+\pi^-$.
We combine pairs of oppositely charged tracks extrapolated to a 
common vertex to form $\phi$ and \KS candidates.
For the charged tracks from the $\phi$ decay we require at least 12 
measured drift-chamber (DCH) coordinates and a minimal transverse momentum \pt of 
0.1~\gevc. The tracks must also originate within 1.5~cm in $xy$ and 
$\pm 10$~cm along the $z$-axis of the nominal beam spot.
Tracks used to reconstruct the $\phi$ mesons are distinguished from pion
and proton tracks via a requirement on a likelihood ratio 
that combines \dedx information 
from the silicon vertex tracker (SVT) and the DCH
for tracks with momentum $p < 0.7 \gevc$. 
For tracks with higher $p$, \dedx in the DCH and the 
Cherenkov angle and the number of photons as measured by the 
internally reflecting ring-imaging Cherenkov detector are used
in the likelihood.
The two-kaon invariant mass must be within 16~\mevcc of the 
nominal $\phi$ mass~\cite{pdg}. 

For tracks corresponding to \KS and $B_{\rm tag}$ daughters
our requirements are less restrictive.
A $\KS\rightarrow\pi^+\pi^-$ candidate is accepted if its
two-pion invariant mass is within 15~\mevcc of the nominal $K^0$ mass~\cite{pdg}, 
its reconstructed decay vertex is separated from the collision point
by at least 3 standard deviations, and the angle 
between the line connecting the $\phi$ and \KS\ decay 
vertices and the \KS momentum direction is less than 45~mrad.

We identify a \KL candidate as in our $\Bz\to J/\psi \KL$ analysis~\cite{macfprd}
either as a cluster of energy deposits in the electromagnetic calorimeter (EMC) 
or as a cluster of hits in two or more layers of the instrumented flux return (IFR) 
that cannot be associated with any charged track in the event. 
The \KL energy is not well measured. Therefore, we determine the \KL
laboratory momentum from its flight direction as measured from the EMC or IFR
cluster and the constraint that the invariant $\phi\KL$ mass agrees with the
known $B^0$ mass. In those cases where the \KL is detected in both the IFR and EMC 
we use the angular information from the EMC, as it has a higher precision.
In order to reduce background from $\pi^0$ decays, we reject
an EMC \KL\ candidate cluster if it forms an invariant mass between 100
and 150~\mevcc with any other cluster in the event under the $\gamma\gamma$ 
hypothesis, or if it has energy greater than 1~GeV and contains two shower 
maxima consistent with two photons from a $\pi^0$ decay.  
The remaining background of photons and overlapping showers is further reduced 
with the use of a neural network constructed from cluster shape variables, 
trained on Monte Carlo (MC) simulated $B^0\to\phi\KL$ and measured radiative 
Bhabha events, and tested on measured $e^+e^-\to\phi(\to\KS\KL )\gamma$ and 
$B^0\to J/\psi \KL$ events. The final $\phi\KL$ sample consists of 
approximately equal numbers of IFR and EMC \KL\ candidates.

The results are extracted from an extended unbinned maximum likelihood 
fit for which we parameterize the distributions of several
kinematic and topological variables for signal and background events
in terms of probability density functions (PDFs)~\cite{oldpub}.
The background arises primarily from random combinations of tracks 
produced in events of the type $e^+e^-\to q\bar{q}$, where  
$q = u,d,s,c$ (continuum). 
Background from other $B$ decay final states 
with and without charm is estimated with MC simulations.
Opposite \CP contributions from the $K^+K^-K^0$ final state 
(e.g. $K^+K^-$ S-wave) are estimated with an angular analysis 
on data to be less than 6.6\% and treated as systematic error. 
The shapes of event variable distributions are obtained from signal 
and background MC samples and high statistics data control samples.

Each $B_{CP}$ candidate is characterized by the energy difference 
$\Delta E = E_B^* - \frac{1}{2}\sqrt{s}$ and, except for $\Bz\to\phi\KL$,
the beam-energy--substituted mass 
$\mes = \sqrt{(\frac{1}{2}s + \vec{p}_0\cdot\vec{p}_B)^2/E_0^2 - p_B^2}$~\cite{Aubert:2001tu}.
The subscripts 0 and $B$ refer to the initial $\Upsilon(4S)$ and the $B_{CP}$ candidate,
respectively, and the asterisk denotes the $\Upsilon(4S)$ rest frame.
For signal events, \DeltaE is expected to peak at zero and 
\mes at the nominal $B$ mass. 
We require $\Delta E < 0.08$~GeV for $\Bz\to\phi\KL$ and
$|\DeltaE|<0.2$~GeV and $\mes > 5.2 \gevcc$ for $\Bz\to\phi\KS$.
In the fit we also use the helicity angle $\theta_H$, which is defined 
as the angle between the directions of the $K^+$ and the parent $B_{CP}$
in the $K^+K^-$ rest frame. The $\cos\theta_H$ distribution 
for pseudoscalar-vector $B$ decay modes is $\cos^2\theta_H$,
and for the combinatorial background it is nearly uniform. 

In continuum events, particles appear bundled into jets. This topology can 
be characterized with several variables computed in the CM frame.
One such quantity is the angle $\theta_T$ between the thrust axis of the 
$B_{CP}$ candidate and the thrust axis formed from the other charged and 
neutral particles in the event. 
We also use the angle $\theta_B$ between the $B_{CP}$ momentum and 
the beam axis, and the sum of the momenta $p_i$ of the other charged and neutral 
particles in the event weighted by the Legendre polynomials $L_n(\theta_i), n=0,2$,
where $\theta_i$ is the angle between the momentum of particle $i$
and the thrust axis of the $B_{CP}$ candidate.
For $\Bz\to\phi\KS$ candidates, we combine these variables 
into a Fisher discriminant ${\mathcal F}$~\cite{Fisher:et}
after requiring $|\cos\theta_T| < 0.9$.
In this mode background from other $B$ decays is negligible, as 
demonstrated in MC simulation studies. 

More stringent criteria must be applied to supress backgrounds 
in the case of $\Bz\to\phi\KL$ candidates, 
and we require $|\cos\theta_T| < 0.8$ and $|\cos\theta_B|<0.85$.
We define the missing momentum $\vec{p}_{miss}$,
calculated in the laboratory frame from the beam momentum
and all tracks and EMC clusters, excluding the \KL candidate.  
We require the polar angle $\theta_{miss}$ of the
missing momentum with respect to the beam direction 
to be greater than 0.3~rad. The cosine of the
angle between $\vec{p}_{miss}$ and the \KL direction, $\theta_K$, must
satisfy $\cos\theta_K > 0.6$.  In the plane transverse to the beam
direction, the difference between the missing momentum projected along the
\KL direction and the calculated \KL momentum
must be greater than $-0.75$~\gevc. In the Fisher discriminant
we replace $|\cos\theta_B|$ by the cosine of the angle 
between the missing momentum and the $K^+$ from the  $\phi$ decay.
In the $\phi\KL$ sample about 1.4\% of the events originate from 
charm $B$ decays,
0.7\% originate from charmless $B$ decays, and about 0.2\% potentially have a 
\CP asymmetry. The dominant contamination is the mode $B\to\phi K^*$, where 
the $K^*$ decays to $\KL\pi$. In the likelihood fit we explicitly parameterize 
backgrounds from both charm and charmless $B$ decays as derived from MC 
simulations.
\par
All the other tracks and clusters in the event are used to form the $B_{\text{tag}}$,
and its flavor is determined with a multivariate tagging algorithm~\cite{sin2bnewbabar}. 
The tagging efficiency $\epsilon$ and mistag probability $w$ in four hierarchical and 
mutually exclusive categories is measured from fully reconstructed 
$B^0$ decays into the $D^{(*)-}X^+\,(X^+ = \pip, \rho^+, a_1^+)$  and 
$\jpsi K^{*0}\,(K^{*0}\to\Kp\pim)$ flavor eigenstates (\Bflav sample).
The analyzing power $\epsilon (1-2w)^2$ is $(28.7\pm 0.7)$\%. 
\par
A detailed description of the $\Delta t$ reconstruction algorithm is 
given in Ref.~\cite{macfprd}.
The $B_{CP}$ vertex resolution is dominated by the $\phi$ vertex. 
The average $\deltaz$ resolution is $190\mum$ 
and is dominated by the tagging vertex in the event.
Thus, we can characterize the resolution with the much larger 
$B_{\text{flav}}$ sample, which we fit simultaneously with the \CP samples.
The amplitudes for the $B_{CP}$ asymmetries and for the $B_{\text{flav}}$ 
flavor oscillations are reduced by the same factor due to 
wrong tags. Both distributions are convoluted with a common $\Delta t$ 
resolution function, and the backgrounds are accounted for by adding
terms to the likelihood, incorporated with different assumptions 
about their $\Delta t$ evolution and 
resolution function~\cite{macfprd}.

\begin{figure}[ht]
\begin{center}
\begin{tabular}{ll}
\epsfig{file=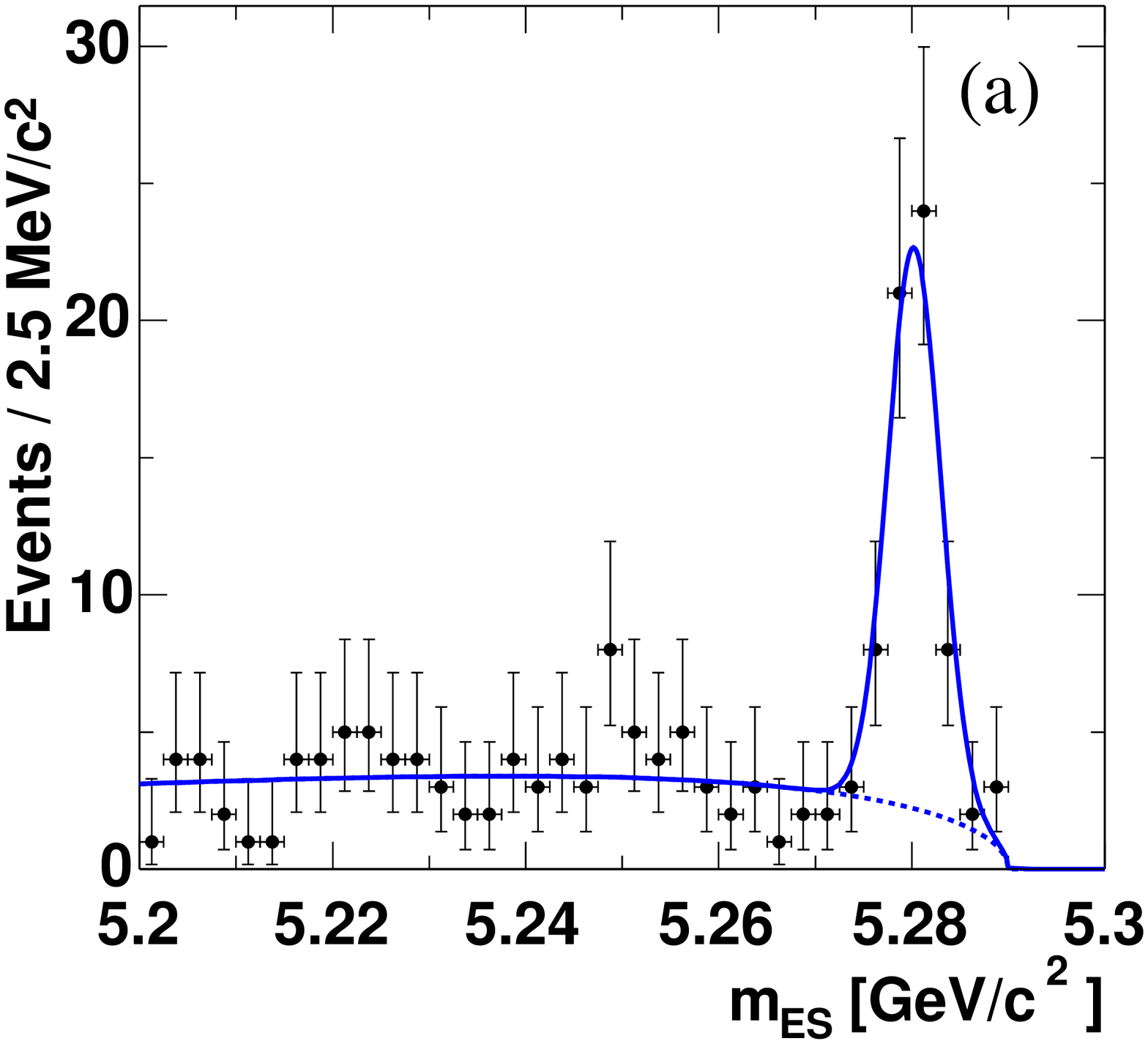,width=4.35cm} 
\epsfig{file=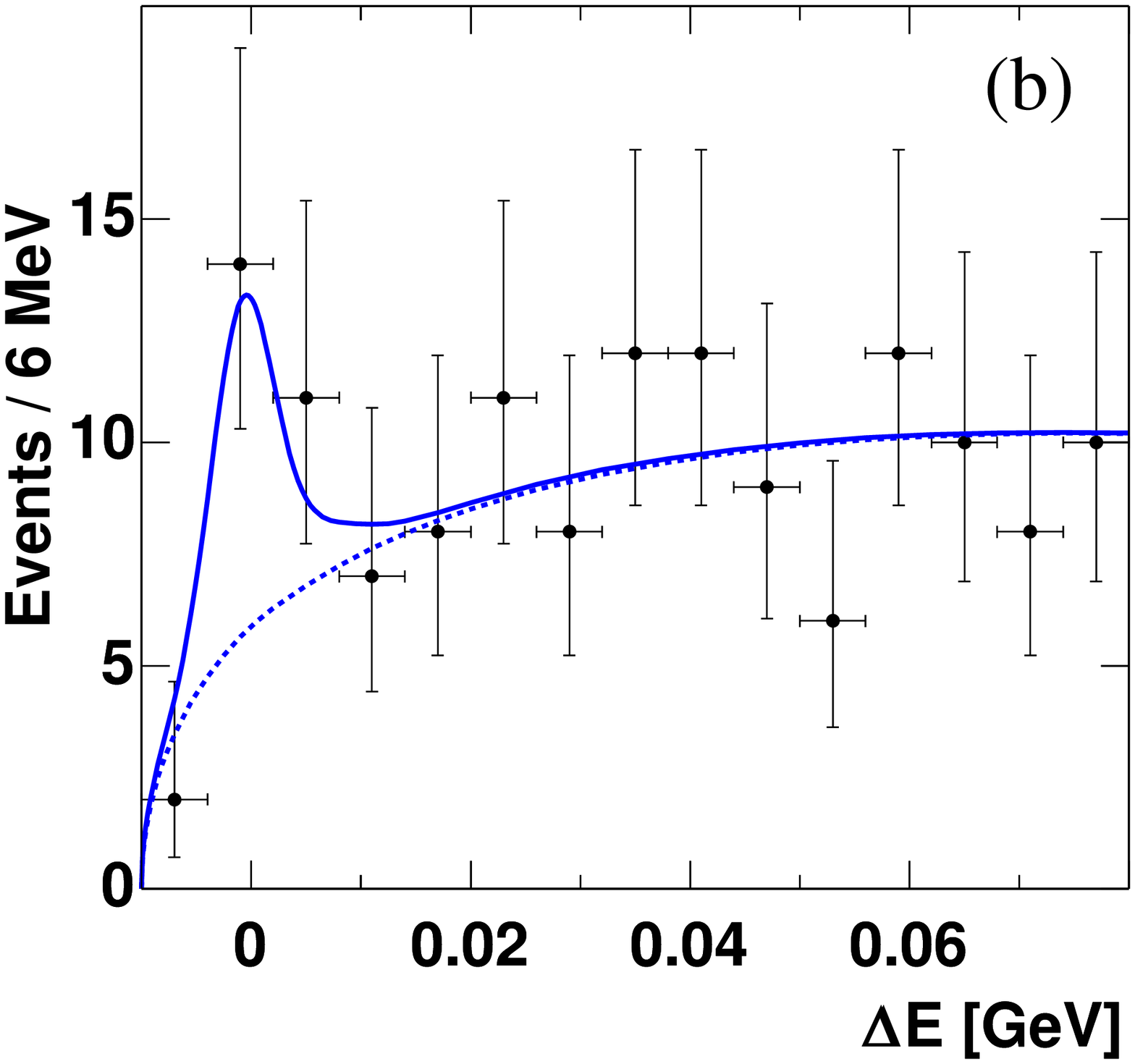,width=4.35cm}
\end{tabular} 
\caption{Distribution of the event variable (a) $m_{ES}$ for the $\phi \KS$ final state
and (b) $\Delta E$ for the $\phi \KL$ final state after reconstruction and
a requirement for the likelihood with total signal efficiency of
32\% and 5\%, respectively. The solid line represents the fit result
for the total event yield and the dotted line for the background.
\label{fig:yield}}
\end{center}
\end{figure}

Since we measure the correlations among the observables to be small in
the data samples entering the fit, we take the probability density
function $\mathcal{P}_{i,c}^j$ for each event $j$ to be a product of the
PDFs for the separate observables. For each event hypothesis $i$
(signal, background) and tagging category $c$, we define
$\mathcal{P}_{i,c}^j =\mathcal{P}_i(m_{ES})\cdot\mathcal{P}_i(\Delta E)
\cdot\mathcal{P}_i(\mathcal{F})
\cdot\mathcal{P}_i(\cos\theta_H)
\cdot\mathcal{P}_i(\Delta t;\sigma_{\Delta t}, c)$,
where for the $\phi\KL$ mode $\mathcal{P}_i(m_{ES}) = 1$ and for the
flavor sample $\mathcal{P}_i(\mathcal{F})\cdot\mathcal{P}_i(\cos\theta_H) = 1$.
The $\sigma_{\Delta t}$ is the error on $\Delta t$ for a given event.
The likelihood function for each decay chain is then
\begin{equation}
{\mathcal L} = \prod_{c}\exp{\left(-\sum_{i}N_{i,c}\right)}
\prod_{j}^{N_c}\left[\sum_{i}N_{i,c}\,{\mathcal P}_{i,c}^j \right] ,
\end{equation}
where $N_{i,c}$ is the yield of events of hypothesis $i$ found by the
fitter in category $c$, and $N_c$ is the number of category $c$ events in the sample.
The total sample consists of 86200 $B_{\text{flav}}$, 
2138 $\phi\KS$ and 4730 $\phi\KL$ candidates. We find $70\pm 9$ $\phi\KS$ and 
$52\pm 16$ $\phi\KL$ signal events.
The signal yields in both the $\phi K^0$ channels
agree well with our determination of the branching fraction 
for $B^0\to\phi K^0$~\cite{sasha}.
Fig.~\ref{fig:yield} shows the \mes (\DeltaE ) distribution
of $\phi \KS$ ($\phi \KL$) events together with the result from the fit 
after a requirement on the likelihood (computed without the variable
plotted) to enhance the sensitivity.

We determine the $\CP$ parameters $\spk$ and $\cpk$ along with an 
additional 38 free parameters:
the efficiency per tagging category (4 parameters),
the average mistag fraction and the difference  
between \Bz and \Bzb mistags for each tagging category (8 parameters), 
the signal \deltat resolution (9), and time dependence (6), 
\deltat resolution (3) and mistag fractions (8) for the background.
We fix $\tau_{\Bz}$ and $\deltamd$ to the world averages~\cite{pdg}. 
The determination of the mistag fractions and \deltat-resolution
parameters is dominated by the high-statistics \Bflav sample. 
The fit was tested with a parameterized simulation of a large
number of data-sized experiments and full detector simulated events
for the different signal and background samples. 
The fit was also verified with our $J/\psi \KS$ data sample and
a control sample of 232 $\phi K^+$ candidates
where one expects \spkp = \cpkp = 0. 
We measure $\spkp = 0.23\pm 0.24$ and $\cpkp = -0.14\pm 0.18$
with statistical errors only. 
The simultaneous fit to the $\phi K^0$ and flavor decay modes yields:
\begin{eqnarray}
\spk & = & 0.47\pm 0.34\mbox{(stat)}\, ^{+0.08}_{-0.06}\mbox{(syst)}, \nonumber \\
\cpk & = & 0.01\pm 0.33\mbox{(stat)} \pm 0.10\mbox{(syst)}. \nonumber
\end{eqnarray}
The result in the dominant channel $B^0\to\phi\KS$ is
$\spk = 0.45\pm 0.43$ and $\cpk = -0.38\pm 0.37$ with statistical errors only.
\begin{figure}[t]
\begin{center}
\epsfig{file=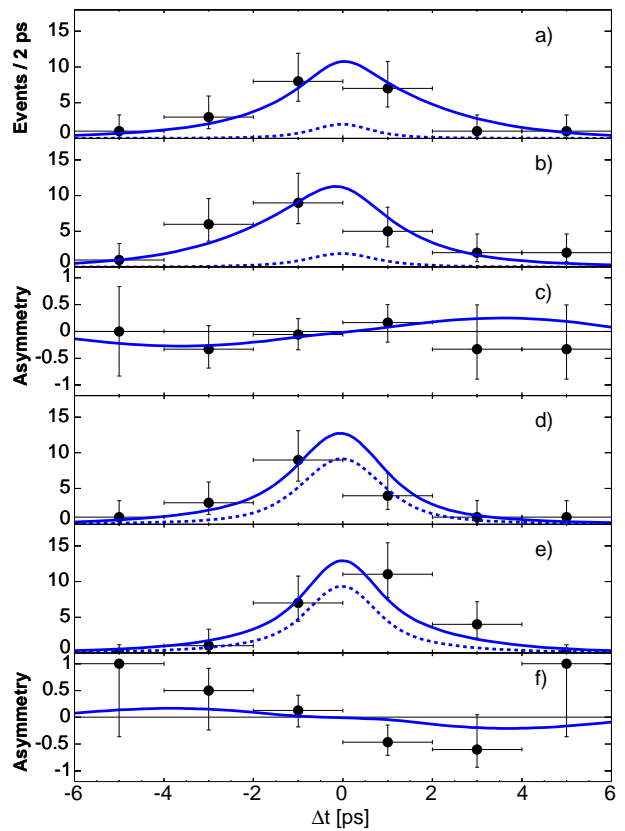, width=8.6cm} 
\caption{Plots a) and b) show the $\Delta t$ distributions of  
\Bz- and \Bzb-tagged $\phi\KS$ events. 
The solid lines refer to the fit for all events; the dashed lines 
correspond to the background. Plot c) shows the asymmetry.
A requirement for the event likelihood is applied.
Plots d), e), and f) are the corresponding plots for $\phi\KL$ 
events.
\label{fig:b0b0bar}}
\vspace*{-0.5cm}
\end{center}
\end{figure}
Fig.~\ref{fig:b0b0bar} shows the $\deltat$ distributions of the \Bz- and the 
\Bzb-tagged subsets together with the raw asymmetry for the $\phi \KS$ 
and $\phi\KL$ events with the result of the combined time-dependent 
\CP-asymmetry fit superimposed.

We consider systematic uncertainties in the \CP coefficients \spk and \cpk 
due to the event-yield determination in the two channels 
(\ppm 0.01 for \spk, \ppm 0.05 for \cpk ),
contributions from \Bz final states with opposite \CP (+0.06, \ppm 0.02),
the parameterization of PDFs for the event yield in signal and 
background (\ppm 0.02, \ppm 0.05), 
composition and \CP asymmetry of the background in the \CP 
events (\ppm 0.03, \ppm 0.03), 
the assumed parameterization of the $\Delta t$ resolution 
function (\ppm 0.02, \ppm 0.01), 
the \mes\ background parameterization (\ppm 0.02, \ppm 0.05),
a possible difference in the efficiency for \Bz and \Bzb (\ppm 0.01, \ppm 0.02),
the fixed values for $\Delta m_d$ and $\tau_B$ (\ppm 0.00, \ppm 0.01),
the beam-spot position (\ppm 0.01, \ppm 0.01), and
uncertainties in the SVT alignment (\ppm 0.01, \ppm 0.01).
The bias in the coefficients due to the fit procedure (\ppm 0.03, \ppm 0.01) is
included in the uncertainty without making corrections to the final results. 
We estimate errors due to the effect of doubly 
CKM-suppressed decays~\cite{Long:2003wq} to be (\ppm 0.01, \ppm 0.03).
We add these contributions in quadrature to obtain the total
systematic uncertainty.

In summary, we have measured the time-dependent \CP asymmetries
in the combined $B$-meson final states $\phi \KS$ and $\phi \KL$.
We obtain values for the \CP -violation parameters \spk and \cpk 
that agree within one standard deviation with the ones measured
in the charmonium channels~\cite{sin2bnewbabar,sin2bnewbelle}; 
the central value of \spk is also consistent with no \CP asymmetry
at the 1.3 $\sigma$ level. 

\begin{acknowledgments}
We are grateful for the excellent luminosity and machine conditions
provided by our \pep2\ colleagues, 
and for the substantial dedicated effort from
the computing organizations that support \babar.
The collaborating institutions wish to thank 
SLAC for its support and kind hospitality. 
This work is supported by
DOE
and NSF (USA),
NSERC (Canada),
IHEP (China),
CEA and
CNRS-IN2P3
(France),
BMBF and DFG
(Germany),
INFN (Italy),
FOM (The Netherlands),
NFR (Norway),
MIST (Russia), and
PPARC (United Kingdom). 
Individuals have received support from the 
A.~P.~Sloan Foundation, 
Research Corporation,
and Alexander von Humboldt Foundation.
\end{acknowledgments}

\bibliographystyle{h-physrev2-original}   

\end{document}